\documentstyle[preprint,aps,epsfig,floats]{revtex}

\begin{document}

\def\eref#1{(\ref{#1})}
\def\av#1{\left\langle#1\right\rangle}
\def\vv{{\bf v}}
\def\vM{{\bf M}}
\def\O{{\rm O}}
\def\e{{\rm e}}
\def\i{{\rm i}}
\def\d{{\rm d}}
\def\half{\mbox{$\frac12$}}

\newdimen\captwidth
\captwidth=5.5in
\def\capt#1{\refstepcounter{figure}\bigskip\hbox to \textwidth{%
       \hfil\vbox{\hsize=\captwidth\renewcommand{\baselinestretch}{1}\small
       {\sc Figure \thefigure}\quad#1}\hfil}\bigskip}

\title{Exact solution of site and bond percolation\\
on small-world networks}
\author{Cristopher Moore$^{1,2}$ and M. E. J. Newman$^2$}
\address{$^1$Departments of Computer Science and Physics, University of
New Mexico,\\
Albuquerque, New Mexico 87131}
\address{$^2$Santa Fe Institute, 1399 Hyde Park Road, Santa Fe,
New Mexico 87501}
\maketitle

\begin{abstract}  
  We study percolation on small-world networks, which has been proposed as
  a simple model of the propagation of disease.  The occupation
  probabilities of sites and bonds correspond to the susceptibility of
  individuals to the disease and the transmissibility of the disease
  respectively.  We give an exact solution of the model for both site and
  bond percolation, including the position of the percolation transition at
  which epidemic behavior sets in, the values of the two critical exponents
  governing this transition, and the mean and variance of the distribution
  of cluster sizes (disease outbreaks) below the transition.
\end{abstract}

\newpage

In the late 1960s, Milgram performed a number of experiments which led him
to conclude that, despite there being several billion human beings in the
world, any two of them could be connected by only a short chain of
intermediate acquaintances of typical length about six~\cite{Milgram67}.
This result, known as the ``small-world effect'', has been confirmed by
subsequent studies and is now widely believed to be correct, although
opinions differ about whether six is an accurate estimate of typical chain
length~\cite{Watts99}.

The small-world effect can be easily understood in terms of random
graphs~\cite{Bollobas85} for which typical vertex--vertex distances
increase only as the logarithm of the total number of vertices.  However,
random graphs are a poor representation of the structure of real social
networks, which show a ``clustering'' effect in which there is an increased
probability of two people being acquainted if they have another
acquaintance in common.  This clustering is absent in random graphs.
Recently, Watts and Strogatz~\cite{WS98} have proposed a new model of
social networks which possesses both short vertex--vertex distances and a
high degree of clustering.  In this model, sites are arranged on a
one-dimensional lattice of size $L$, and each site is connected to its
nearest neighbors up to some fixed range $k$.  Then additional
links---``shortcuts''---are added between randomly selected pairs of sites
with probability $\phi$ per link on the underlying lattice, giving an
average of $\phi k L$ shortcuts in total.  The short-range connections
produce the clustering effect while the long-range ones give average
distances which increase logarithmically with system size, even for quite
small values of $\phi$.

This model, commonly referred to as the ``small-world model,'' has
attracted a great deal of attention from the physics community.  A number
of authors have looked at the distribution of path lengths in the model,
including scaling forms~\cite{BA99,NW99b,Moukarzel99} and exact and
mean-field results~\cite{KAS00,NMW99}, while others have looked at a
variety of dynamical systems on small-world
networks~\cite{WS98,Monasson99,BW00}.  A review of recent developments can
be found in Ref.~\onlinecite{Newman00}.

One of the most important consequences of the small-world effect is in the
propagation of disease.  Clearly a disease can spread much faster through a
network in which the typical person-to-person distance is $\O(\log L)$ than
it can through one in which the distance is $\O(L)$.
Epidemiology recognizes two basic parameters governing the effects of a
disease: the {\em susceptibility}---the probability that an individual
exposed to a disease will contract it---and the {\em
  transmissibility}---the probability that contact between an infected
individual and a healthy but susceptible one will result in the latter
contracting the disease.  Newman and Watts~\cite{NW99b} studied a model of
disease in a small-world which incorporates these variables.  In this model
a randomly chosen fraction $p$ of the sites or bonds in the small-world
model are ``occupied'' to represent the effects of these two parameters.  A
disease outbreak which starts with a single individual can then spread only
within a connected cluster of occupied sites or bonds.  Thus the problem of
disease spread maps onto a site or bond percolation problem.  At some
threshold value $p_c$ of the percolation probability, the system undergoes
a percolation transition which corresponds to the onset of epidemic
behavior for the disease in question.  Newman and Watts gave an approximate
solution for the position of this transition on a small-world network.

In this paper, we give an exact solution for both site and bond percolation
on small-world networks using a generating function method.  Our method
gives not only the exact position of the percolation threshold, but also
the values of the two critical exponents governing behavior close to the
transition, the complete distribution of the sizes of disease outbreaks for
any value of $p$ below the transition, and closed-form expressions for the
mean and variance of the distribution.  A calculation of the value of $p_c$
only, using a transfer-matrix method, has appeared previously as
Ref.~\onlinecite{MN99}.

The basic idea behind our solution is to find the distribution of ``local
clusters''---clusters of occupied sites or bonds on the underlying
lattice---and then calculate how the shortcuts join these local clusters
together to form larger ones.  We focus on the quantity $P(n)$, which is
the probability that a randomly chosen site belongs to a connected cluster
of $n$ sites.  This is also the probability that a disease outbreak
starting with a randomly chosen individual will affect $n$ people.  It is
{\em not\/} the same as the distribution of cluster sizes for the
percolation problem, since the probability of an outbreak starting in a
cluster of size $n$ increases with cluster size in proportion to~$n$, all
other things being equal.  The cluster size distribution is therefore
proportional to $P(n)/n$, and can be calculated easily from the results
given in this paper, although we will not do so.

\begin{figure}[t]
\begin{center}
\psfig{figure=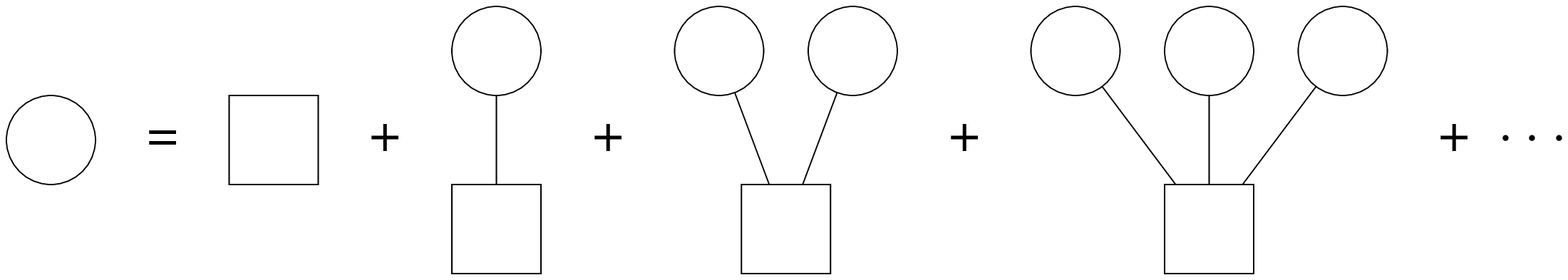,width=5.5in}
\end{center}
\capt{Graphical representation of a cluster of connected sites.  The entire
  cluster (circle) is equal to a single local cluster (square), with any
  number $m\ge0$ of complete clusters attached to it by a single shortcut.}
\label{sum}
\end{figure}

We start by examining the site percolation problem, which is the simpler
case.  Since $P(n)$ is difficult to evaluate directly, we turn to a
generating function method for its calculation.  We define
\begin{equation}
H(z) = \sum_{n=0}^\infty P(n) z^n.
\end{equation}
For all $p<1$, as we show below, the distribution of local clusters falls
off with cluster size exponentially, so that every shortcut leads to a
different local cluster for $L$ large: the probability of two shortcuts
connecting the same pair of local clusters falls off as $L^{-1}$.  This
means that any complete cluster of sites consists of a local cluster with
$m\ge0$ shortcuts leading from it to $m$ other clusters.  Thus $H(z)$
satisfies the Dyson-equation-like iterative condition illustrated
graphically in Fig.~\ref{sum}, and we can write it self-consistently as
\begin{equation}
H(z) = \sum_{n=0}^\infty P_0(n) z^n \sum_{m=0}^\infty P(m|n) [H(z)]^m.
\label{iter1}
\end{equation}
In this equation $P_0(n)$ is the probability of a randomly chosen site
belonging to a local cluster of size $n$, which is
\begin{equation}
P_0(n) = \Biggl\lbrace \begin{array}{ll}
            1-p                 & \qquad\mbox{for $n=0$}\\
            n p q^{n-1} (1-q)^2 & \qquad\mbox{for $n\ge1$,}
          \end{array}
\label{p0}
\end{equation}
with $q=1-(1-p)^k$.  $P(m|n)$ is the probability of there being exactly $m$
shortcuts emerging from a local cluster of size $n$.  Since there are
$2\phi k L$ ends of shortcuts in the network, $P(m|n)$ is given by the
binomial
\begin{equation}
P(m|n) = \biggl({2\phi k L\atop m}\biggr) \biggl[{n\over L}\biggr]^m
         \biggl[1-{n\over L}\biggr]^{2\phi k L-m}.
\label{pln}
\end{equation}
Using this expression Eq.~\eref{iter1} becomes
\begin{equation}
H(z) = \sum_{n=0}^\infty P_0(n) z^n
         \biggl[ 1 + \bigl(H(z)-1\bigr){n\over L}\biggr]^{2\phi k L}
     = \sum_{n=0}^\infty P_0(n) \bigl[ z \e^{2k\phi(H(z)-1)} \bigr]^n,
\label{iter2}
\end{equation}
for $L$ large.  The remaining sum over $n$ can now be performed
conveniently by defining
\begin{equation}
H_0(z) = \sum_{n=0}^\infty P_0(n) z^n = 1 - p + pz{(1-q)^2\over(1-qz)^2},
\label{h0}
\end{equation}
where the second equality holds in the limit of large $L$ and we have made
use of~\eref{p0}.  $H_0(z)$ is the generating function for the local
clusters.  Now we notice that $H(z)$ in Eq.~\eref{iter2} is equal to
$H_0(z)$ with $z\to z \e^{2k\phi(H(z)-1)}$.  Thus
\begin{equation}
H(z) = H_0\bigl(z \e^{2k\phi(H(z)-1)}\bigr).
\label{iter3}
\end{equation}

$H(z)$ can be calculated directly by iteration of this equation starting
with $H(z)=1$ to give the complete distribution of sizes of epidemics in
the model.  It takes $n$ steps of the iteration to calculate $P(n)$
exactly.  The first few steps give
\begin{eqnarray}
P(0) &=& 1 - p,\\
P(1) &=& p (1-q)^2 \e^{-2k\phi p},\\
P(2) &=& p (1-q)^2 \bigl[ 2q + 2k\phi p (1-q)^2 \bigr] \e^{-4k\phi p}.
\end{eqnarray}
It is straightforward to verify that these are correct.  We could also
iterate Eq.~\eref{iter3} numerically and then estimate $P(n)$ using, for
instance, forward differences at $z=0$.  Unfortunately, like many
calculations involving numerical derivatives, this method suffers from
severe machine-precision problems which limit us to small values of $n$, on
the order of $n\lesssim20$.  A much better technique is to evaluate $H(z)$
around a contour in the complex plane and calculate the derivatives using
the Cauchy integral formula:
\begin{equation}
P(n) = {1\over n!}\,{\d^n\!H\over\d z^n}\bigg|_{z=0}
     = {1\over2\pi\i} \oint {H(z)\over z^{n+1}} \d z.
\label{cauchy}
\end{equation}
A good choice of contour in the present case is the unit circle $|z|=1$.
Using this method we have been able to calculate the first thousand
derivatives of $H(z)$ without difficulty.

\begin{figure}[t]
\begin{center}
\psfig{figure=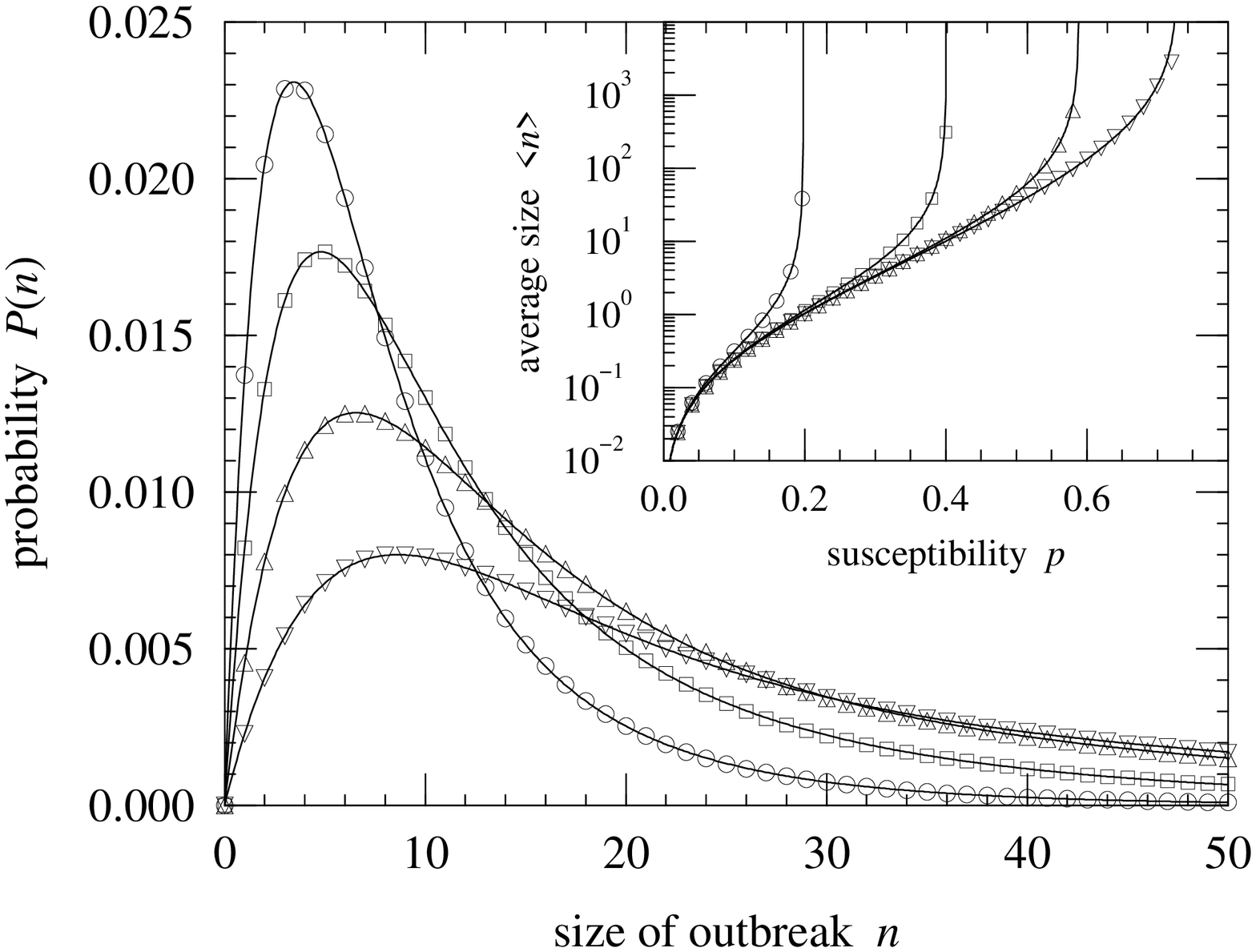,width=4.5in}
\end{center}
\capt{The distribution of outbreak sizes in simulations of the site
  percolation model with $L=10^7$, $k=5$, $\phi=0.01$, and $p=0.25$,
  $0.30$, $0.35$, and $p=p_c=0.40101$ (circles, squares, and up- and
  down-pointed triangles respectively).  The solid lines are the same
  distributions calculated using Eqs.~\eref{iter3} and~\eref{cauchy}.
  Inset: The average size of disease outbreaks as a function of $p$ for
  (left to right) $\phi=10^{-1}, 10^{-2}, 10^{-3}, 10^{-4}$.  The points
  are numerical results for $L=10^7$, $k=5$ and the solid lines are the
  exact result, Eq.~\eref{avn}.}
\label{outbreak}
\end{figure}

In Fig.~\ref{outbreak} we show the distribution of outbreak sizes as a
function of $n$ calculated from Eq.~\eref{cauchy} for a variety of values
of $p$.  On the same plot we also show the distribution of outbreaks
measured in computer simulations of the model on systems of $L=10^7$ sites.
As the figure shows, the agreement between the two is excellent.

We can also calculate any moment of the distribution in closed form using
Eq.~\eref{iter3}.  For example, the mean outbreak size is given by the
first derivative of $H$:
\begin{equation}
\av{n} = H'(1) = {H_0'(1)\over1-2k\phi H_0'(1)}
       = {p(1+q)\over1-q-2k\phi p(1+q)},
\label{avn}
\end{equation}
and the variance is given by
\begin{eqnarray}
\langle n^2 \rangle - \av{n}^2 &=& H''(1) + H'(1) - [{H'}(1)]^2,\nonumber\\
 &=& {p [1 + 3q - 3q^2 - q^3 - p(1-q)(1+q)^2
     + 2k\phi p^2 (1+q)^3 ]\over[1 - q - 2k\phi p(1+q)]^3}.
\end{eqnarray}

In the inset of Fig.~\ref{outbreak} we show Eq.~\eref{avn} for various
values of $\phi$ along with numerical results from simulations of the
model, and the two are again in good agreement.

\begin{figure}[t]
\begin{center}
\psfig{figure=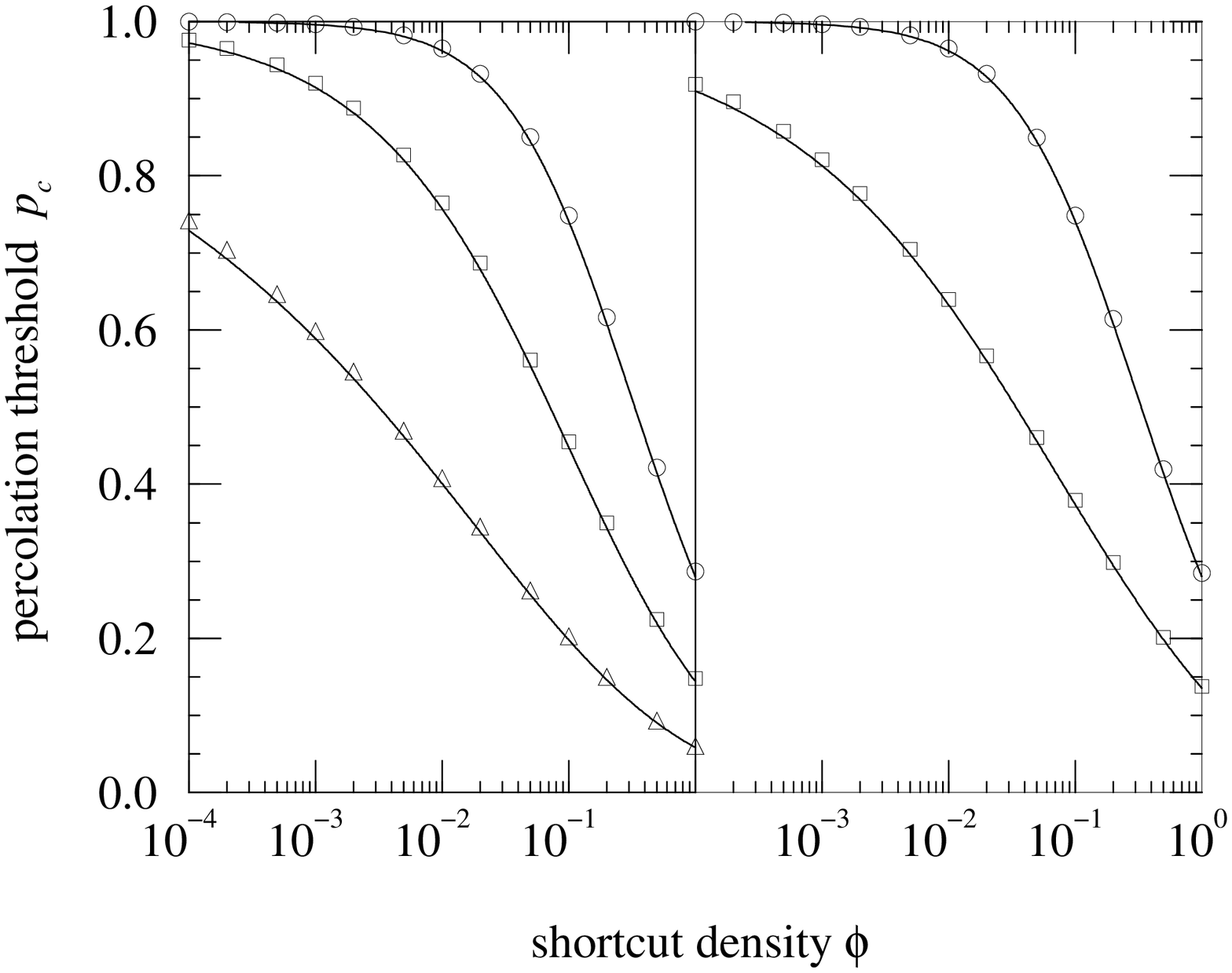,width=4.5in}
\end{center}
\capt{Numerical results for the percolation threshold as a function of
  shortcut density~$\phi$ for systems of size $L=10^6$ (points).  Left
  panel: site percolation with $k=1$ (circles), $2$~(squares), and
  $5$~(triangles).  Right panel: bond percolation with $k=1$~(circles) and
  $2$~(squares).  The solid lines are the analytic expressions for the same
  quantities, Eqs.~\eref{sitethresh}, \eref{bondthreshk1},
  and~\eref{bondthreshk2}.}
\label{perc}
\end{figure}

The mean outbreak size diverges at the percolation threshold $p=p_c$.  This
threshold marks the onset of epidemic behavior in the model~\cite{NW99b}
and occurs at the zero of the denominator of Eq.~\eref{avn}.  The value of
$p_c$ is thus given by
\begin{equation}
\phi = {1-q_c\over2kp_c(1+q_c)} = {(1-p_c)^k\over2kp_c(2-(1-p_c)^k)},
\label{sitethresh}
\end{equation}
in agreement with Ref.~\onlinecite{MN99}.  The value of $p_c$ calculated
from this expression is shown in the left panel of Fig.~\ref{perc} for
three different values of $k$.

The denominator of Eq.~\eref{avn} is analytic at $p=p_c$ and has a non-zero
first derivative with respect to $p$, so that to leading order the
divergence in $\av{n}$ goes as $(p_c-p)^{-1}$ as we approach percolation.
Defining a critical exponent $\sigma$ in the conventional fashion $\av{n}
\sim (p_c-p)^{-1/\sigma}$, we then have
\begin{equation}
\sigma = 1.
\end{equation}
Near $p_c$ we expect $P(n)$ to behave as
\begin{equation}
P(n) \sim n^{-\tau} \e^{-n/n^*}\qquad\mbox{as $n\to\infty$.}
\label{tail}
\end{equation}
It is straightforward to show~\cite{HL14} that both the typical outbreak
size $n^*$ and the exponent $\tau$ are governed by the singularity of
$H(z)$ closest to the origin: $n^* = (\log z^*)^{-1}$, where $z^*$ is the
position of the singularity, and $\tau=\alpha+1$, where
$H(z)\sim(z^*-z)^\alpha$ close to $z^*$.

In general, the singularity of interest may be either finite or not; the
order of the lowest derivative of $H(z)$ which diverges at $z^*$ depends on
the value of $\alpha$.  In the present case, $H(z^*)$ is finite but the
first derivative diverges, and we can use this to find $z^*$ and $\alpha$.

Although we do not have a closed-form expression for $H(z)$, it is simple
to derive one for its functional inverse $H^{-1}(w)$.  Putting $H(z)\to w$
and $z\to H^{-1}(w)$ in Eq.~\eref{iter3} and rearranging we find
\begin{equation}
H^{-1}(w) = H_0^{-1}(w)\,\e^{2k\phi(1-w)}.
\label{inverse}
\end{equation}
The singularity in $H(z)$ corresponds to the point $w^*$ at which the
derivative of $H^{-1}(w)$ is zero, which gives $2k\phi z^* H_0'(z^*)=1$,
making $z^*=\e^{1/n^*}$ a real root of the cubic equation
\begin{equation}
(1-qz)^3 - 2k\phi pz (1-q)^2 (1+qz) = 0.
\end{equation}
The second derivative of $H^{-1}(w)$ is non-zero at $w^*$, which implies
that $H(z)\sim(z^*-z)^{1/2}$ and hence $\alpha=\frac12$ and the outbreak
size exponent is
\begin{equation}
\tau = \mbox{$\frac32$}.
\label{valuetau}
\end{equation}
A power-law fit to the simulation data for $P(n)$ shown in
Fig.~\ref{outbreak} gives $\tau = 1.501\pm0.001$ in good agreement with
this result.

The values $\sigma=1$ and $\tau=\frac32$ put the small-world percolation
problem in the same universality class as percolation on a random
graph~\cite{SA92}, which seems reasonable since the effective dimension of
the small-world model in the limit of large system size is
infinite~\cite{NW99b} just as it is for a random graph.

We close our analysis of the site percolation problem by noting that
Eq.~\eref{iter3} is similar in structure to the equation $H(z) = ze^{H(z)}$
for the generating function of the set of rooted, labeled trees.  This
leads us to conjecture that it may be possible to find a closed-form
expression for the coefficients of the generating function $H(z)$ using the
Lagrange inversion formula~\cite{Wilf94}.

Turning to bond percolation, we can apply the same formalism as above with
only two modifications.  First, the probability $P_0(n)$ that a site
belongs to a local cluster of size $n$ is different for bond percolation
and consequently so is $H_0(z)$ (Eq.~\eref{h0}).  For the case $k=1$
\begin{equation}
P_0(n) = n p^{n-1} (1-p)^2,
\end{equation}
where $p$ is now the bond occupation probability.  This expression is the
same as Eq.~\eref{p0} for the site percolation case except that $P_0(0)$ is
now zero and $P_0(n\ge1)$ contains one less factor of $p$.  $H_0(z)$ for
$k=1$ is
\begin{equation}
H_0(z) = z {(1-p)^2\over(1-pz)^2}.
\end{equation}
For $k>1$, calculating $P_0(n)$ is considerably more complex, and in fact
it is not clear whether a closed-form solution exists.  However, it is
possible to write down the form of $H_0(z)$ directly using the method given
in Ref.~\onlinecite{MN99}.  For $k=2$, for instance,
\begin{equation}
H_0(z) = {z (1 - p)^4 \bigl(1 - 2 p z + p^3 (1 - z) z + p^2 z^2
         \bigr)\over1 - 4 p z + p^5 (2 - 3z) z^2 - p^6 (1 - z) z^2
         + p^4 z^2 (1 + 3z) + p^2 z (4 + 3z) - p^3 z \bigl(1 + 5z + z^2\bigr)}.
\end{equation}
The second modification to the method is that in order to connect two local
clusters a shortcut now must not only be present (which happens with
probability $\phi$) but must also be occupied (which happens with
probability $p$).  This means that every former occurrence of $\phi$ is
replaced with $\phi p$.  The rest of the analysis follows through as before
and we find that $H(z)$ satisfies the recurrence relation
\begin{equation}
H(z) = H_0\bigl(z \e^{2k\phi p(H(z)-1)}\bigr),
\label{iter4}
\end{equation}
with $H_0$ as above.  Thus, for example, the mean outbreak size is now
\begin{equation}
\av{n} = H'(1) = {H_0'(1)\over1-2k\phi p H_0'(1)},
\end{equation}
and the percolation transition occurs at $2k\phi p H_0'(1) = 1$, which
gives
\begin{equation}
\phi = {1-p_c\over2 p_c (1+p_c)}
\label{bondthreshk1}
\end{equation}
for $k=1$ and
\begin{equation}
\phi = {(1 - p_c)^3 (1 - p_c + p_c^2)\over4 p_c (1 + 3 p_c^2 - 3 p_c^3
       - 2 p_c^4 + 5 p_c^5 - 2 p_c^6)}
\label{bondthreshk2}
\end{equation}
for $k=2$.  As in the site percolation case, the critical exponents are
$\sigma=1$ and $\tau=\frac32$.  In the right panel of Fig.~\ref{perc} we
show curves of $p_c$ as a function of $\phi$ for the bond percolation model
for $k=1$ and $k=2$, along with numerical results for the same quantities.
The agreement between the exact solution and the simulation results is
good.

We can also apply our method to the case of simultaneous site and bond
percolation, by replacing $P_0(n)$ with the appropriate distribution of
local cluster sizes and making the replacement $\phi\to\phi p_{\rm bond}$
as above.  The developments are simple for the case $k=1$ but the
combinatorics become tedious for larger $k$ and so we leave these
calculations to the interested (and ambitious) reader.

To conclude, we have studied the site and bond percolation problems in the
Watts--Strogatz small-world model as a simple model of the spread of
disease.  Using a generating function method we have calculated exactly the
position of the percolation transition at which epidemics first appear, the
values of the two critical exponents describing this transition, and the
sizes of disease outbreaks below the transition.  We have confirmed our
results with extensive computer simulations of disease spread in
small-world networks.

Finally, we would like to point out that the method described here can in
principle be extended to small-world networks built on underlying lattices
of higher dimensions~\cite{Watts99,NW99b}.  Only the generating function
for the local clusters $H_0(z)$ needs to be recalculated, although this is
no trivial task; such a calculation for a square lattice with $k=1$ would
be equivalent to a solution of the normal percolation problem on such a
lattice, something which has not yet been achieved.  Even without a
knowledge of $H_0(z)$, however, it is possible to deduce some results.  For
example, we believe that the critical exponents will take the values
$\sigma=1$ and $\tau=\frac32$, just as in the one-dimensional case, for the
exact same reasons.  It would be possible to test this conjecture
numerically.

\vspace{7mm}
{\small The authors are grateful to Michael Renardy for pointing out
Eq.~\eref{inverse}, and to Keith Briggs, Noam Elkies, Philippe Flajolet,
and David Rusin for useful comments.  This work was supported in part by
the Santa Fe Institute and DARPA under grant number ONR
N00014--95--1--0975.}

\end{document}